\documentclass[twocolumn,secnumarabic,amssymb,amsmath,nofootinbib,tightenlines,nobibnotes,aps,prl,showpacs]{revtex4}
\usepackage{graphicx,epsfig}

\begin{document}

\title{Universal Pion Freeze-out in Heavy-Ion Collisions}

\def\rez{$^{(1)}$}
\def\gsi{$^{(2)}$}
\def\hei{$^{(3)}$}
\def\dub{$^{(4)}$}
\def\wei{$^{(5)}$}
\def\sun{$^{(6)}$}
\def\cer{$^{(7)}$}
\def\bnl{$^{(8)}$}
\def\mpi{$^{(9)}$}

\author{D.~Adamov\'a\rez, 
G.~Agakichiev\gsi, 
H.~Appelsh\"auser\hei, 
V.~Belaga\dub, 
P.~Braun-Munzinger\gsi, 
A.~Castillo\gsi,
A.~Cherlin\wei, 
S.~Damjanovi\'c\hei, 
T.~Dietel\hei, 
L.~Dietrich\hei, 
A.~Drees\sun, 
S.\,I.~Esumi\hei, 
K.~Filimonov\hei, 
K.~Fomenko\dub,
Z.~Fraenkel\wei, 
C.~Garabatos\gsi, 
P.~Gl\"assel\hei, 
G.~Hering\gsi, 
J.~Holeczek\gsi, 
V.~Kushpil\rez, 
B.~Lenkeit\cer, 
W.~Ludolphs\hei,
A.~Maas\gsi, 
A.~Mar\'{\i}n\gsi, 
J.~Milo\v{s}evi\'c\hei,
A.~Milov\wei, 
D.~Mi\'skowiec\gsi, 
Yu.~Panebrattsev\dub, 
O.~Petchenova\dub, 
V.~Petr\'a\v{c}ek\hei, 
A.~Pfeiffer\cer, 
J.~Rak\mpi, 
I.~Ravinovich\wei, 
P.~Rehak\bnl, 
H.~Sako\gsi, 
W.~Schmitz\hei, 
J.~Schukraft\cer, 
S.~Sedykh\gsi, 
S.~Shimansky\dub, 
J.~Sl\'{\i}vov\'a\hei,
H.\,J.~Specht\hei, 
J.~Stachel\hei, 
M.~\v{S}umbera\rez, 
H.~Tilsner\hei, 
I.~Tserruya\wei, 
J.\,P.~Wessels\gsi, 
T.~Wienold\hei, 
B.~Windelband\hei, 
J.\,P.~Wurm\mpi, 
W.~Xie\wei, 
S.~Yurevich\hei, 
V.~Yurevich\dub \\
W.~Schmitz\hei, 
J.~Schukraft\cer, 
S.~Sedykh\gsi, 
S.~Shimansky\dub, 
J.~Sl\'{\i}vov\'a\hei,
H.\,J.~Specht\hei, 
J.~Stachel\hei, 
M.~\v{S}umbera\rez, 
H.~Tilsner\hei, 
I.~Tserruya\wei, 
J.\,P.~Wessels\gsi, 
T.~Wienold\hei, 
B.~Windelband\hei, 
J.\,P.~Wurm\mpi, 
W.~Xie\wei, 
S.~Yurevich\hei, 
V.~Yurevich\dub \\
(CERES Collaboration) }

\address{
\rez NPI ASCR, \v{R}e\v{z}, Czech Republic\\
\gsi GSI Darmstadt, Germany\\
\hei Heidelberg University, Germany\\
\dub JINR Dubna, Russia\\
\wei Weizmann Institute, Rehovot, Israel\\
\sun SUNY at Stony Brook, U.S.A.\\
\cer CERN, Geneva, Switzerland\\
\bnl BNL, Upton, U.S.A.\\
\mpi MPI, Heidelberg, Germany\\
}

\begin{abstract}
Based on an evaluation of data on pion interferometry and on particle
yields at mid-rapidity, we propose a universal condition for thermal
freeze-out of pions in heavy-ion collisions. We show that freeze-out
occurs when the mean free path of pions $\lambda_{f}$ reaches a value
of about 1~fm, which is much smaller than the spatial extent of the
system at freeze-out.  This critical mean free path is independent of
the centrality of the collision and beam energy from AGS to RHIC.

\end{abstract}

\maketitle 

A systematic study of the space-time extent and the dynamical behavior
of the pion source in relativistic heavy ion collisions at thermal
freeze-out can be obtained from analysis of pion interferometry (HBT)
data. Understanding these aspects is vital for interpretation of
the data in terms of formation of the quark-gluon plasma. Indeed
recent HBT results from RHIC and how they fit into the systematics have
been noted as a major puzzle \cite{larry}. In this letter we present
an investigation of the freeze-out conditions at beam energies from
AGS to RHIC. In particular, the recently published CERES HBT data at
40, 80, and 158 A GeV~\cite{cereshbt} provide an important link
between the existing results from AGS, SPS, and RHIC, thereby shedding
light on the RHIC puzzle.

Thermal freeze-out of pions and its connection to the mean free
path has been discussed previously (see e.g.~\cite{nagamiya,stock,vischer,bodrum,johannaqm93,akkelin}).
The mean free path of pions at freeze-out is defined as 
\begin{equation}
	\lambda_{f}=\frac{1}{\rho_f \cdot \sigma}=\frac{V_f}{N \cdot \sigma}, 
\label{mfp2}
\end{equation}
where $\sigma$ is the total cross
section of pions with the surrounding medium and $\rho_f$ is the
freeze-out density which can be replaced by the number of particles $N$ contained in
the freeze-out volume $V_f$.

The pion freeze-out volume $V_f$ can be accessed experimentally by
pion interferometry. Mid-rapidity pion HBT data have been published
from central collisions of lead and gold nuclei over a wide range of
beam energies. Here, we focus on recent HBT results from three
experiments which have kinematical access to the region of low
transverse pair momentum
$k_t$$=$$\frac{1}{2}|\vec{p}_{t,1}+\vec{p}_{t,2}|$: Experiment E895 at
the AGS~\cite{E895hbt}, the CERES/NA45 experiment at the
SPS~\cite{cereshbt,heinzphd}, and the STAR experiment at
RHIC~\cite{starhbt}. All three experiments employ large volume Time
Projection Chambers (TPCs), thereby applying comparable analysis
methods with similar sources of systematic uncertainties.

For the calculation of the freeze-out volume $V_f$ we use the following 
expression:
\begin{equation}
  V_f = (2\pi)^{\frac{3}{2}}\cdot R_{\rm long}\cdot R_{\rm side}^{2},
  \label{volume}
\end{equation}
assuming a density distribution of Gaussian shape in all three
dimensions. The longitudinal and sideward radius parameters $R_{\rm
long}$ and $R_{\rm side}$ are measured in the longitudinal co-moving
system of the pion pair, using the cartesian decomposition
of the three-momentum difference vector $\vec{q}$=$(q_{\rm
long},q_{\rm side},q_{\rm out})$ as proposed in \cite{ber}.

The definition of a freeze-out volume in heavy ion collisions has to
be taken with some care because of the strong collective expansion of
the system. Collective expansion leads to space-momentum correlations
of the emitted pions, resulting in a $k_t$-dependent reduction of the
observed HBT radius parameters as compared to the `true' geometric
dimensions of the source~\cite{pratt1,maksin,sinyu,prattcso}.  We
prefer to use a definition containing only measured quantities rather
than model dependent parameters that may not be appropriate at all
beam energies.  But to avoid strong bias of the extracted beam energy
dependence due to expansion we have selected and compared data
measured at similar $\langle k_t \rangle$ values of about 0.16~GeV/c:
The E895 data were taken at $\langle k_t \rangle$$=$0.148~GeV/c, while
the STAR data are at $\langle k_t \rangle$$=$0.170~GeV/c. From the
CERES data we have calculated the average of the results in the
$k_t$-bins at $\langle k_t \rangle$$=$0.125~GeV/c and $\langle k_t
\rangle$$=$0.195~GeV/c.

At all beam energies under investigation, the observed strong decrease
of $R_{\rm long}$ as function of $k_t$ is consistent with a
boost-invariant expansion in longitudinal direction~\cite{maksin}.  In
this limiting case, the geometric size of the system in longitudinal
direction is much larger than the measured homogeneity length $R_{\rm
long}$. The corresponding homogeneity scale in velocity space is given
by the average thermal velocity $\langle \beta_{th} \rangle$ of the
pions.  The thermal velocity $\langle \beta_{th} \rangle$ can be
calculated relativistically using the expression:
\begin{equation}
\langle \gamma_{th} \rangle -1 = \frac{1}{3}\left(\frac{K_1(m_t/T_f)}{K_2(m_t/T_f)}-1\right)+
\frac{T_f}{m_t},
\label{vtherm}
\end{equation}
with $\langle \gamma_{th} \rangle$=$(1-\langle \beta_{th}
\rangle^2)^{-\frac{1}{2}}$, $m_t$$=$$(m_{\pi}^2+k_t^2)^\frac{1}{2}$,
the thermal freeze-out temperature $T_f$, and Bessel functions $K_1$
and $K_2$.  Using Eq.(\ref{vtherm}), we obtain $\langle \beta_{th}
\rangle$=0.7 at $k_t$$=$0.16~GeV/c, assuming $T_f$$=$120~MeV.  Turning
this into a rapidity results in $y_{th}$=${\rm arctanh}(0.7)$=0.87.
We conclude that a source element of longitudinal size $R_{\rm long}$
corresponds to 0.87 units of rapidity at $k_t =$0.16~GeV/c. Note that
$R_{\rm long}$ is the rms of the length of homogeneity and so is the
corresponding width in rapidity. As for the spatial distribution we
assume a Gaussian shape also in rapidity.

Also the transverse radius parameter $R_{\rm side}$ is reduced in the
presence of collective transverse flow. Model dependent
approximations~\cite{chapnix,csolo,scheibl} predict a reduction by
about 20-25\% as compared to the true geometric transverse source size
at $k_t$$=$0.16~GeV/c for $T_f$$=$120~MeV and a typical average
transverse flow velocity $\langle v_t
\rangle$$=$0.5$c$~\cite{cereshbt,na49hbt,borisana}. We prefer not to
use this reduction in the analysis but rather include the
corresponding model dependent underestimate of the volume in the
systematic errors below. The observed beam energy dependence of $T_f$
and $\langle v_t \rangle$~\cite{xuqm} imposes an additional systematic
uncertainty. Within reasonable limits of $T_f$ and $\langle v_t
\rangle$, the relative change of $R_{\rm side}$ however is small and
even partially compensated, as both $T_f$ and $\langle v_t \rangle$
increase with beam energy.

\begin{figure}[t!]
\begin{center}
   \includegraphics[bbllx=120,bblly=10,bburx=400,bbury=520,width=7.5cm]{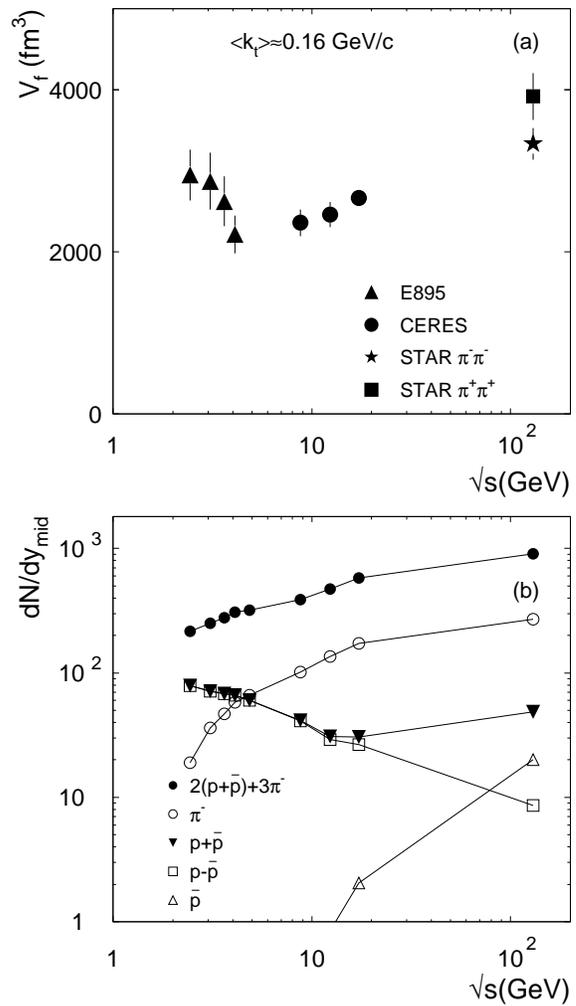}
   \caption{(a): The freeze-out volume $V_f$ as function of
   $\sqrt{s}$. (b): Mid-rapidity densities of negative pions, protons,
   and antiprotons vs.~$\sqrt{s}$.}
\label{fig1}
\end{center}
\end{figure}

The experimentally determined freeze-out volume $V_f$, calculated
according to Eq.(\ref{volume}) as function of $\sqrt{s}$, is shown in
Fig.~\ref{fig1}(a). We observe a steep decrease of $V_f$ at AGS
energies and an increase throughout the SPS energy regime towards
RHIC.  The data indicate the existence of a minimum between AGS and
SPS energies. The origin of this non-monotonic behaviour
cannot be understood in terms of previously presented freeze-out
scenarios, where constant particle density in coordinate space or
constant pion phase space density have been proposed as possible
universal freeze-out conditions~\cite{pom,ferenc}.  This is
demonstrated in Fig.~\ref{fig1}(b), where a compilation of the
mid-rapidity densities d$N$/d$y$$\left.\right|_{y_{\rm mid}}$ of pions
and protons as function of $\sqrt{s}$ is shown
~\cite{cebra,seyboth,phenixyields,e895prot,e917prot,na49prot,na44prot}.
The number of pions and the sum of pions and nucleons are
monotonically increasing with $\sqrt{s}$, without indication of a
minimum.
 
For the determination of the critical mean free path $\lambda_f$, we
need to evaluate the denominator in Eq.(\ref{mfp2}). In the presence
of different particle species~$i$ we replace $N$$\cdot$$\sigma$ by a
sum:
\begin{equation}
	N\cdot \sigma = \sum_{i} N_i \sigma_{\pi i} = N_N \cdot
	\sigma_{\pi N} + N_\pi \cdot \sigma_{\pi\pi}. 
\end{equation}
For simplicity we neglect less abundant particle species such as
kaons and deuterons. Following our previous argument we integrate over
the rapidity distribution with an rms of $y_{th}$=0.87 and account for
the different isospin states, leading to: 
\begin{eqnarray}
N_N&=&2\cdot \sqrt{2\pi} \cdot 0.87\cdot {\rm d}N_{p+\bar{p}}/{\rm d}y\left.\right|_{y_{\rm mid}}, ~~{\rm and} \nonumber \\
N_\pi&=&3\cdot \sqrt{2\pi} \cdot 0.87\cdot {\rm d}N_{\pi^-}/{\rm d}y\left.\right|_{y_{\rm mid}}.
\end{eqnarray} 
For the cross sections we use $\sigma_{\pi\pi}$$=$13~mb, and a thermal
and isospin averaged value $\sigma_{\pi N}$$=$72~mb
\cite{johannaqm93}. The latter assumes a thermal freeze-out
temperature $T_f$=120~MeV independent of beam energy
and ignoring possible changes due to the
non-equilibrium nature of thermal freeze-out.

\begin{figure}[t!]
\begin{center}
   \includegraphics[bbllx=120,bblly=10,bburx=400,bbury=520,width=7.5cm]{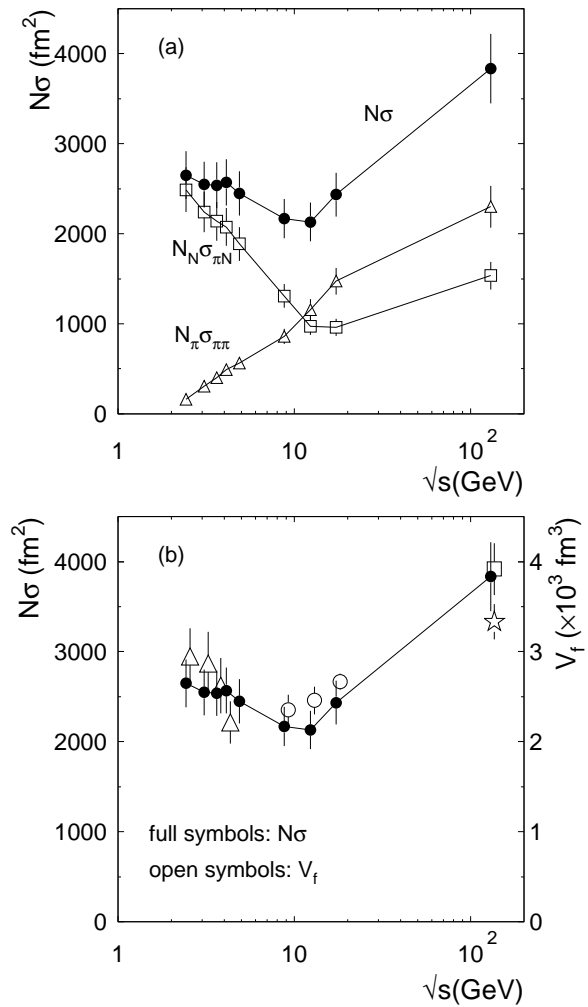}
   \caption{(a): Beam energy dependence of $N\cdot \sigma$ (for 
	explanation see text). (b): Comparison of $N\cdot \sigma$
	and $V_f$.}
   \label{fig2}
\end{center}
\end{figure}

In Fig.~\ref{fig2}(a) $N$$\cdot$$\sigma$ is shown as function of
$\sqrt{s}$.  At AGS energies $N$$\cdot$$\sigma$ is dominated by the
contribution from nucleons while at RHIC the contribution from pions
exceeds that from nucleons, due to the change in chemical fireball
composition.  It is the large ratio of $\sigma_{\pi N}$/$\sigma_{\pi
\pi}$ combined with the steep increase in pion multiplicity with beam
energy that produces a non-monotonic beam energy dependence of
$N$$\cdot$$\sigma$, leading to a minimum in the transition region
between AGS and SPS energies, in striking similarity with the observed
behaviour of $V_f$ in Fig.~\ref{fig1}(a).  To demonstrate this
quantitatively we have superimposed $V_f$ and $N$$\cdot$$\sigma$ in
Fig.~\ref{fig2}(b). According to Eq.(\ref{mfp2}) we conclude that
thermal pion freeze-out occurs at constant mean free path of
$\lambda_f \approx$ 1.0~fm. We note that $\lambda_f$ is much smaller
than the system size: Thermal freeze-out is determined by the product
of cross section times local density. A similar local pion freeze-out
criterium was discussed in~\cite{bodrum,johannaqm93} albeit only for
fireballs with comparable nucleon and pion numbers. This behaviour
indicates a significant degree of opaqueness of the pion 
source~\cite{vischer}, as it
arises naturally for a collectively expanding system~\cite{borisflow}.
Considering systematic effects on the evaluation of $\lambda_f$ such
as the usage of $R_{\rm side}$ for the volume, the Gaussian assumption
for the spatial and rapidity distributions, the uncertainty in
temperature (assumed as 120 MeV) and thereby in cross sections, and
the uncertainty in evaluating the average thermal velocity, we find
that the first two completely dominate, leading to $0.7 \leq \lambda_f
\leq 1.4 $ fm.
\begin{figure}[t!]
\begin{center}
   \includegraphics[bbllx=120,bblly=10,bburx=400,bbury=520,width=7.5cm]{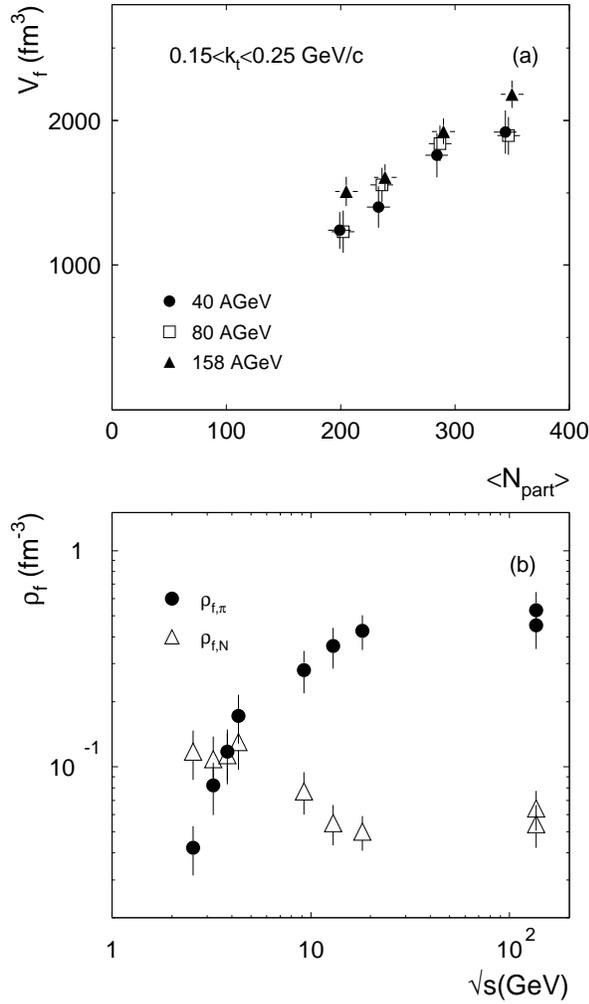}
   \caption{(a): Freeze-out volume as function of $\langle N_{\rm
   part} \rangle$ in Pb+Au collisions (from~\cite{cereshbt}). (b):
   Pion and nucleon freeze-out densities $\rho_{f,\pi}$ and
   $\rho_{f,N}$ as function of $\sqrt{s}$. At $\sqrt{s}$$=$130~GeV,
   results from $\pi^+\pi^+$ and $\pi^-\pi^-$ interferometry are
   shown.}  \label{fig3}
\end{center}
\end{figure}

Pion freeze-out at constant $\lambda_f$ implies that the freeze-out
density $\rho_f$ is constant, as long as the chemical composition of
the fireball does not change. This is indeed demonstrated by recent
results of the CERES collaboration for the freeze-out volume $V_f$ in
40, 80, and 158 AGeV Pb+Au collisions~\cite{cereshbt} shown in
Fig.~\ref{fig3}(a).  Since, at fixed energy, particle
abundances scale approximately linearly with the
number of participating nucleons
$\langle
N_{\rm part} \rangle$~\cite{na45mult,wa98mult,wa97mult}, the
observation of a linear increase of $V_f$ with centrality is
consistent with freeze-out at constant density.

However, as function of beam energy, relative particle abundances do
change, and therefore the freeze-out density $\rho_f$ is in general
not constant. In Fig.~\ref{fig3}(b) we show the pion and nucleon
freeze-out densities $\rho_{f,\pi}=N_{\pi}/V_f$ and
$\rho_{f,N}$$=$$N_{N}/V_f$ as function of $\sqrt{s}$. The freeze-out
densities change drastically with $\sqrt{s}$ at AGS and SPS energies,
however, only little change is observed from top SPS energy to
RHIC. This indicates that the freeze-out densities reach asymptotic
values at high beam energy.

In conclusion, we have derived a universal condition for pion
freeze-out from pion interferometry data and single particle yields.
Thermal pion freeze-out occurs at a critical mean free path
$\lambda_f$$\approx$1~fm, independent of beam energy.  We observe a
transition from nucleon to pion dominated freeze-out between AGS and
SPS energies, characterized by a minimum of the freeze-out volume
$V_f$.  The existence of this minimum appears as a consequence of the
relative change of particle abundances with beam energy, under
consideration of their different cross sections with pions. In this
picture, the overall weak $\sqrt{s}$ dependence of HBT radii up to
RHIC energies finds a simple interpretation. 
The surprisingly small value of $\lambda_f$ points to a considerable opaqueness
of the source~\cite{vischer,borisflow}.
This opaqueness may be at the root of the small observed values
for $R_{\rm out}$ at RHIC~\cite{vischer,borisflow,padula}.

This work was supported by the German BMBF, the U.S.~DoE, the
Israeli Science Foundation, and the MINERVA Foundation.

\end{document}